\documentclass[review]{elsarticle}

\usepackage[colorlinks=true,linkcolor=blue]{hyperref}
\usepackage{lineno}
\usepackage{multirow}
\modulolinenumbers[1]
\journal{Neural Networks (SI: ADLMBIA)}

\biboptions{semicolon,round,sort,authoryear}
\biboptions{authoryear}

\begin{document}
\begin{frontmatter}
\title{End-to-end semantic segmentation of personalized deep brain structures for non-invasive brain stimulation}
\author[a,b]{Essam A. Rashed}
\ead{essam.rashed@nitech.ac.jp}
\author[a]{Jose Gomez-Tames}
\author[a,c]{Akimasa Hirata}
\address[a]{Department of Electrical and Mechanical Engineering, Nagoya Institute of Technology, Nagoya 466-8555, Japan}
\address[b]{Department of Mathematics, Faculty of Science, Suez Canal University, Ismailia 41522, Egypt}
\address[c]{Center of Biomedical Physics and Information Technology, Nagoya Institute of Technology, Nagoya 466-8555, Japan}

\begin{abstract}

Electro-stimulation or modulation of deep brain regions is commonly used in clinical procedures for the treatment of several nervous system disorders. In particular, transcranial direct current stimulation (tDCS) is widely used as an affordable clinical application that is applied through electrodes attached to the scalp. However, it is difficult to determine the amount and distribution of the electric field (EF) in the different brain regions due to anatomical complexity and high inter-subject variability. Personalized tDCS is an emerging clinical procedure that is used to tolerate electrode montage for accurate targeting. This procedure is guided by computational head models generated from anatomical images such as MRI. Distribution of the EF in segmented head models can be calculated through simulation studies. Therefore, fast, accurate, and feasible segmentation of different brain structures would lead to a better adjustment for customized tDCS studies.

In this study, a single-encoder multi-decoders convolutional neural network is proposed for deep brain segmentation. The proposed architecture is trained to segment seven deep brain structures using T1-weighted MRI. Network generated models are compared with a reference model constructed using a semi-automatic method, and it presents a high matching especially in Thalamus (Dice Coefficient (DC) = 94.70\%), Caudate (DC = 91.98\%) and Putamen (DC = 90.31\%) structures. Electric field distribution during tDCS in generated and reference models matched well each other, suggesting its potential usefulness in clinical practice.

\end{abstract}


\end{frontmatter}


\section{Introduction}

Deep brain electrostimulation (modulation) in a non-invasive manner is a clinical procedure applied in the treatment of neurological disorders. One promissing application is transcranial direct current stimulation (tDCS) where a weak direct current can be used to modulate cortical, sub-cortical, and deeper regions \citep{Dasilva2012headache, Frase2016NPhys, Kim2012BS}. Although tDCS is known as an affordable clinical tool, it still suffers from limitations due to high inter- and intra-subject variability that makes it hard to predict the electric current spread in different brain regions \citep{Datta2011BS, Laakso2015BS, Wiethoff2014BS}. Moreover, tDCS bipolar electrode montages can generate consistent current not only in superficial tissues underneath the electrodes but also consistent and significant currents in deep regions at group-level, in which current spread is influenced by all non-brain, cortical and deep brain tissues \citep{GomezTames2019CN}. However, the capability to customize the tDCS treatment scenarios that generate specific currents for specific neurophysiological impact is difficult \citep{Sadleir2012FP, Datta2012FP}. Several published works have studied the effect of subject/anatomy variability on the neuromodulation effects using tDCS \citep{Lopezalonso2014BS, Tremblay2014BS, Laakso2015BS, GomezTames2019JNE}.

The common practice in tDCS is to use one-fits-all electrode montage. However, personalized tDCS is required to reduce the inter-subject difference effects and to increase the potential effectiveness. Simulation studies using personalized head models generated by a segmentation of anatomical images such as MRI is expected to become common in clinical tDCS for dosage optimization \citep{Sadleir2010neuroimage, Thair2017FN}. A volume conductor model representing the patient under study can provide better understating on how likely is the current pathways and what are the potential brain regions that are likely to be stimulated. However, this approach requires a reliable and instance segmentation of major head tissues to be useful in clinical use. Manual segmentation of all head tissues is known to be a tedious time-consuming process that requires special experience for accurate results. On the other hand, automatic segmentation using MRI is a challenging task as several head tissues are presented in low-contrast in MRI (e.g., blood vessels, dura, and spongy bones). Therefore, it is difficult to be identified using standard intensity-based approaches. This problem can be mitigated when additional anatomical information from CT and/or venogram are available but this means additional patient burden.

Convolutional neural networks are evolving as the state-of-art image segmentation techniques for brain MRI \citep{Bernal2019AIM}. With deep learning network architectures, several automatic features can be observed by the network with no need of prior hand-crafted design \citep{Rashed2019pulse}. Recently, several network designs are presented for brain segmentation \citep{Chen2018neuroimage, Wachinger2018NeuroImage, Jog2019neuroimage, Khalili2019MRI, Dolz2019TMI} and deep brain regions \citep{Roy2019NeuroImage, Kushibar2018MIA, Dolz2018NeuroImage, Ryu2019MRI}. Reviews on brain structure segmentation in MRI can be found in \cite{GonzlezVill2016AIM} with an emphasis on deep learning approaches in \cite{Akkus2017JDI}. Although supervised-based segmentation is known to be time-consuming in feature optimization (training phase), it is of reasonable computation cost in evaluation (testing phase). Moreover, an end-to-end network architecture that evolves convolutional operations is highly robust to data vulnerabilities to some extent. 

Previous studies in brain segmentation using deep neural networks can be categorized based on network architecture into three classes. 1) Batch-based networks, in which images/volumes are divided into 2D/3D patches that are used to derive a pixel/voxel oriented features using local neighborhood, 2) semantic-based networks where the whole image is used as network input, and 3) hybrid-networks where both approaches are fused, with potential use of statistical priors such as atlases. 

In this study, a semantic-based end-to-end network architecture is proposed to handle the deep brain region segmentation. In our previous studies \citep{Rashed2019Neuroimage,Rashed2019ICIP}, we presented the ForkNet architecture for full personalized head segmentation. In that work, ForkNet was trained for the segmentation of 13 head standard tissues using T1-weighted MRI for transcranial magnetic stimulation (TMS) of the brain. Here, we extend the ForkNet design for the segmentation of sub-cortical brain regions and validate the accuracy of automatic segmentation for application to tDCS EF estimation. The accuracy of the segmentation is validated by tDCS-generated EFs that are shaped not only by non-brain tissues and cortical tissues but also significantly by deep brain structures.

\begin{figure*}
\centering
\includegraphics[width=\textwidth]{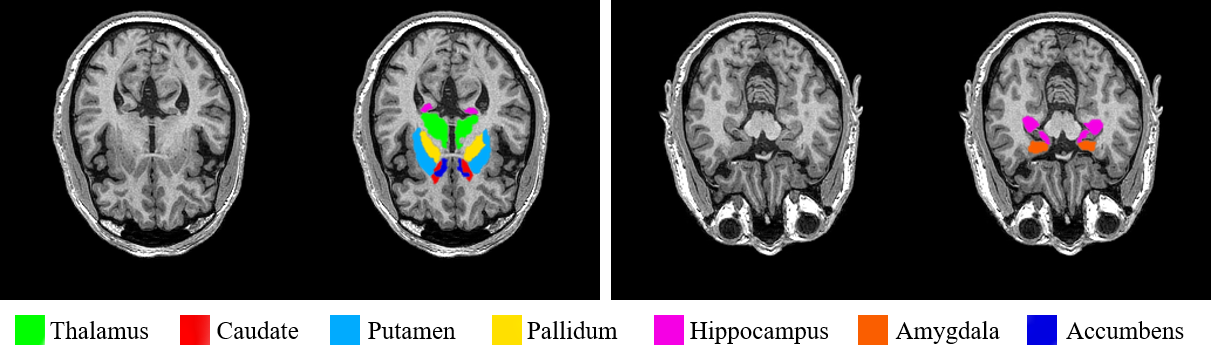}
\caption{An example of deep brain structures segmentation using T1-weighted MRI (NAMIC dataset, subject: case01020). This demonstrate two axial slices with the corresponding segmented deep brain structures overlaid with different colors.} 
\label{example}
\end{figure*}

\section{Materials and methods}

The proposed pipeline to segment deep brain region is presented in this section along with evaluation study using tDCS stimulation. First, we present the datasets used in the study, followed by details of network architecture used in the segmentation process. Next, the segmentation pipeline for deep brain regions is discussed. Finally, we compute tDCS neuromodulation in deep regions using volume conductor models generated using different segmentation methods.

\subsection{MRI Datasets}

In this study, two brain MRI datasets are used for the evaluation of the proposed method. The first dataset is the NAMIC (Brain Multimodality) dataset\footnote{\href{http://hdl.handle.net/1926/1687} {http://hdl.handle.net/1926/1687}} with voxel size 1~mm$^3$ of 18 subjects. The semi-automatic method in our previous study \citep{Laakso2015BS} is used to segment the whole head into different tissues including sub-cortical regions. An example of deep brain region segmentation using the semi-automatic method is shown in Fig.~\ref{example}. The second dataset is the MICCAI 2012 workshop on multi-atlas labeling\footnote{\href{https://my.vanderbilt.edu/masi/workshops/}{https://my.vanderbilt.edu/masi/workshops/}} \citep{Miccai2012} with 35 subjects along with golden truth segmentation of brain structures. Both datasets are used to validate the segmentation accuracy. Segmentation of NAMIC dataset is also used for tDCS studies to validate segmentation quality effect on EF distribution in deep brain structures.

\begin{figure*}
\centering
\includegraphics[width=\textwidth]{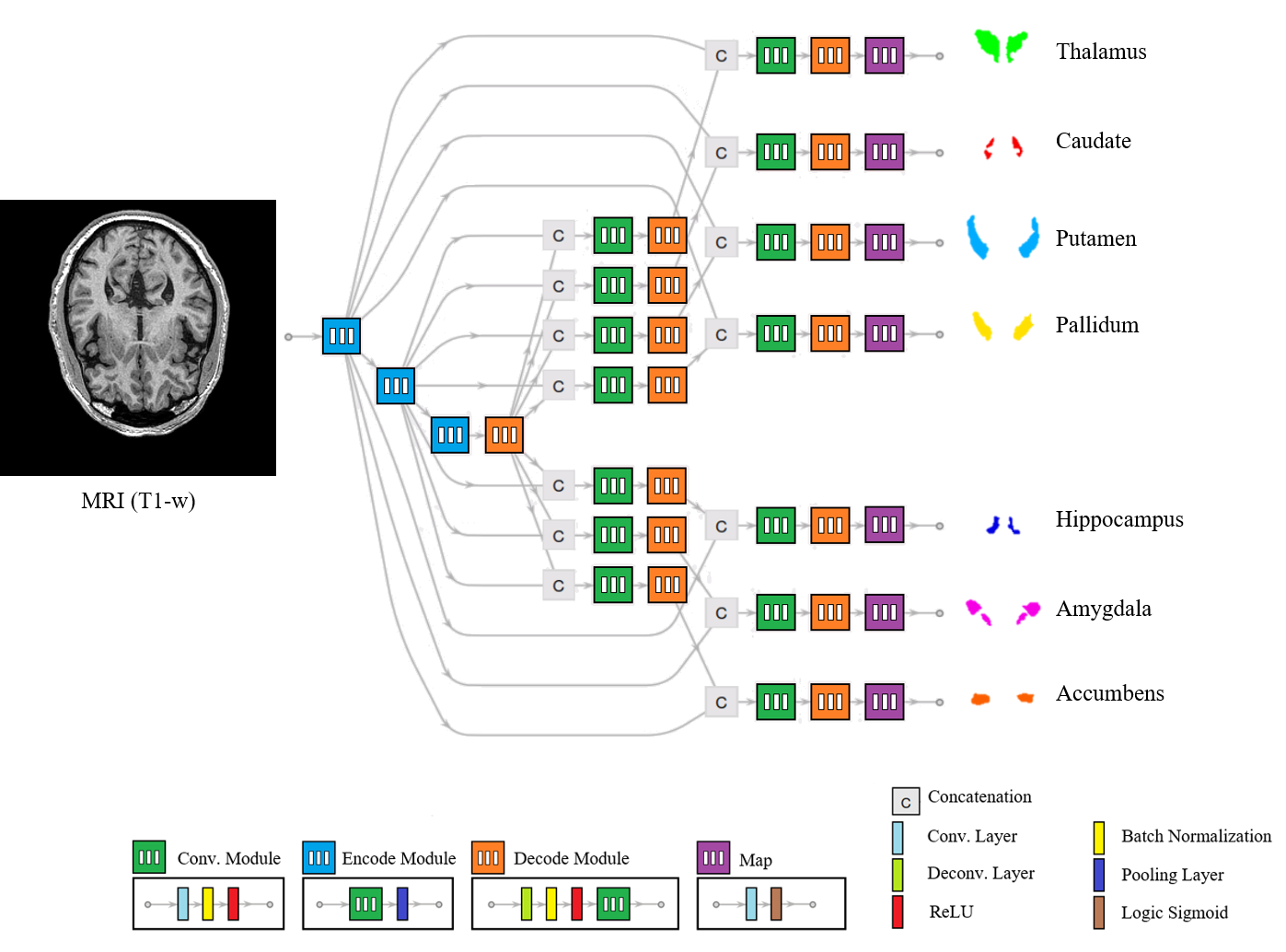}
\caption{The proposed network with degree $N=7$ and depth $D=2$. Network input is a 2D MRI and $N$ outputs are binary labels for different deep brain regions. Details of network architecture are in Table \ref{SubForkSize}.} 
\label{arch}
\end{figure*}

\subsection{Network architecture}

The proposed network architecture is a generalized extension of ForkNet \citep{Rashed2019Neuroimage}. The main design features an end-to-end architecture composite of single track encoders followed by multi-decoders as shown in Fig.~\ref{arch}. The main components are convolution, encoding, decoding, and map modules. Convolution module (ConvMod) consists of a convolutional layer followed by a batch normalization (BN) layer and rectifier linear unit (ReLU) layer. Encode modules (EncMod) are presented in a single track with variable length determined by the network depth $D$. Each encoder is a ConvMod followed by a maximum pooling layer. Decovolution modules (DecMod) are presented in $N$ tracks and $D$ depth. Each deconvolution module is a composition of deconvolution layer, BN layer, ReLU layer, and ConvMod. Finally, each decoding track is attached to a Map modeule which is a convolutional layer and logic sigmoid. This architecture can be customized with degree $N$ identify the number of decoder tracks and depth $D$ refers to how many successive convolutional operations are perfumed. The input is a 2D MRI slice and outputs are probability maps representing different anatomical structures. A key feature of the proposed architecture is the ability to customize decoder tracks individually to fit with texture variability of anatomical structures. Detailed parameters are listed in Table~\ref{SubForkSize}.

Consider $M$ as a volume MRI with $K$ slices, the network output is computed as:

\begin{equation}
L_{k,n}=\textnormal{SubForkNet}(M_k),~k=1,\dots\,K;~ n=1,\dots,N.
\end{equation}

\begin{figure*}
\centering
\includegraphics[width=\textwidth]{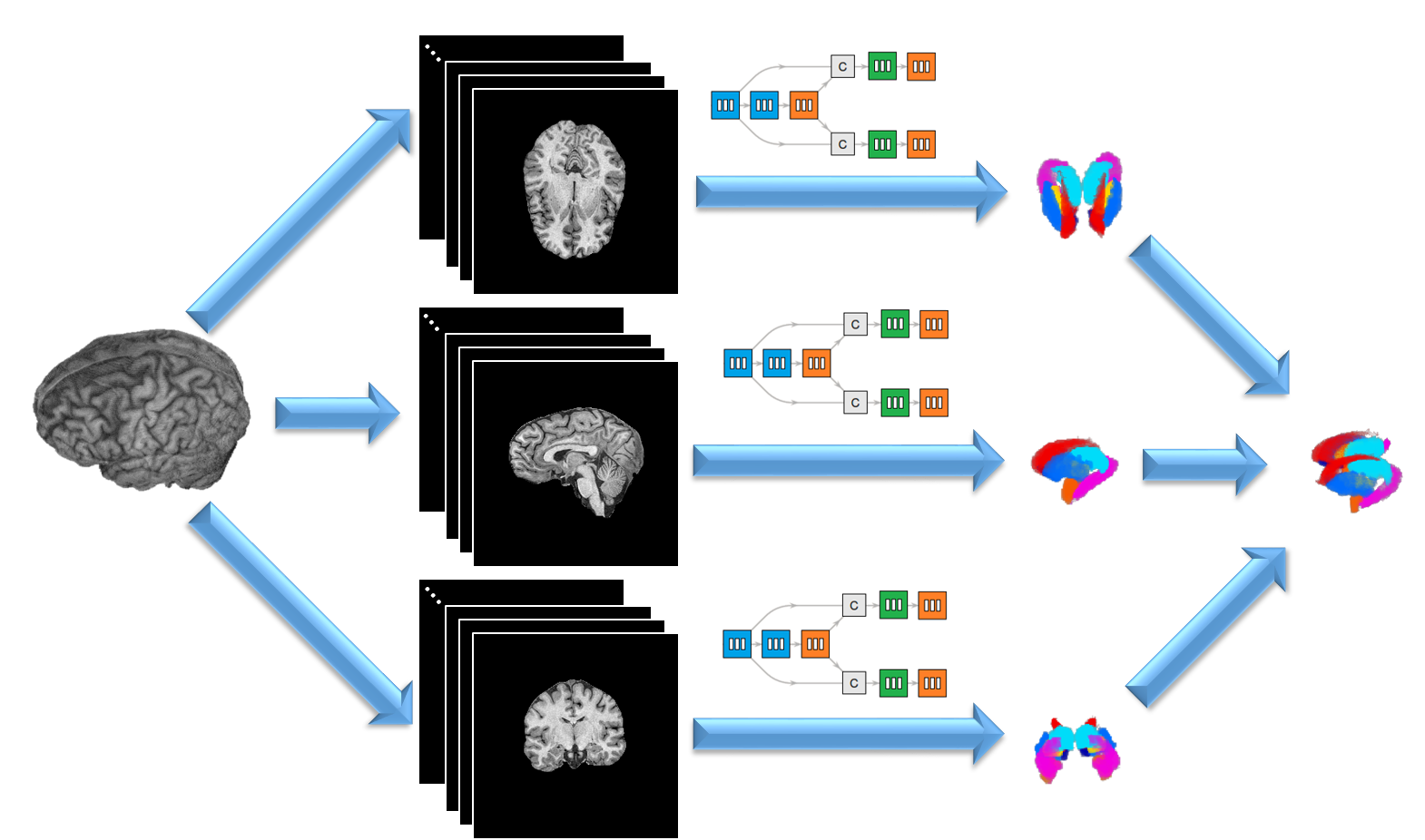}
\caption{Segmentation of deep brain regions using three networks trained using axial (top), sagittal (middle), and coronal (bottom) direction. Aggregation operation is used to compute final segmentation results.} 
\label{model}
\end{figure*}

The corresponding segmented slice is computed using the following SoftMax rule:

\begin{equation}
R_k(i,j)=\left\{\begin{array}{ll} \arg \max_n L_{k,n}(i,j) & \max_n L_{k,n}(i,j)>= \epsilon \\ 0 & \max_n L_{k,n}(i,j)< \epsilon \end{array} \right.
\end{equation}
where $\epsilon$ is a background threshold value. The proposed architecture, named SubForkNet hereafter, is trained using different slicing directions (i.e. axial, sagittal, and coronal) as shown in Fig.~\ref{model}. The rule-based segmentation merge approach using majority vote is used for generate the final segmentation from different slicing directions. When no majority in a voxel is found, the neighborhood majority vote is computed as:

\begin{equation}
R_k^f(i,j)=\arg \max_{t=a,s,c} \max_n \textnormal{Count}_{i,j \in \Omega} R^t_k(i,j), 
\end{equation}
where $R^a, R^s,$ and $R^c$ are segmentation results obtained from axial, sagittal, and coronal directions, respectively and $\Omega$ is a local neighborhood region (here: $\Omega=$3$\times$3).

\begin{table*}
\centering
\footnotesize
\caption{Detailed architecture of SubForkNet (shown in Fig.~\ref{arch}) with degree $N$, Depth $D$, and convolution kernel size $r^2$.}
\begin{tabular}{|ll|lll|l|}
\hline 
{\bf Module} && {\bf layer} & {\bf output size} & {\bf kernel} & {\bf label}\\
\hline \hline
Input & && $2^{8}\times2^{8}$ &&\\
\hline
EncMod$_{i}$ &$i=1 \rightarrow (D+1)$& Convolution & $2^{(i+2)} \times [2^{(9-i)}]^2 $ & $2^{(i+2)}$ $\times r^2$  & \multirow{3}{*}{\includegraphics[width=.5 cm]{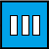}}\\
& &BN \& ReLU & $2^{(i+2)} \times [2^{(9-i)}]^2$ & &   \\
& &Pooling (Max) & $2^{(i+2)} \times [2^{(8-i)}]^2$ && \\
\hline
DecMod$_{j,n}$ &$j=(D+1) \rightarrow 1$& Deconvolution & $2^{(j+1)} \times [2^{(9-j)}]^2 $ & $2^{(j+1)}$ $\times2^2$ &  \multirow{3}{*}{\includegraphics[width=.5 cm]{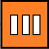}} \\
 & $n=1 \rightarrow N$&BN \& ReLU & $2^{(j+1)} \times [2^{(9-j)}]^2 $ &   & \\
 & &Convolution & $2^{(j+1)} \times [2^{(9-j)}]^2 $ & $2^{(j+1)}$ $\times r^2$  & \\
\hline
ConvMod$_{j,n}$ & $j=D \rightarrow 1$&Convolution & $2^{(j+2)} \times [2^{(8-j)}]^2 $ & $2^{(j+2)}$ $\times r^2$  & \multirow{2}{*}{\includegraphics[width=.5 cm]{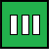}}\\
 & $n=1 \rightarrow N$&BN \& ReLU & $2^{(j+2)} \times [2^{(8-j)}]^2 $ & &   \\
\hline
Concat$_{j,n}$ & $j=D \rightarrow 1$&Concatenation &  $2^{(j+3)} \times [2^{(8-j)}]^2 $ & &\multirow{2}{*}{\includegraphics[width=.4 cm]{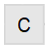}}\\
 &$n=1 \rightarrow N$&  &  & &\\
\hline
Map$_n$ & $n=1 \rightarrow N$&Convolution & $1 \times 2^{8} \times 2^{8}$ & $r^2$&\multirow{2}{*}{\includegraphics[width=.5 cm]{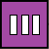}}\\
  && Sigmoid (Log) & $2^{8} \times 2^{8}$  &&\\
\hline
Output$_n$ &$n=1 \rightarrow N$ && $ 2^{8}\times2^{8}$ &&\\
\hline
\end{tabular}
\label{SubForkSize}
\end{table*}

\subsection{Pipeline for personalized head model}

\begin{figure*}
\centering
\includegraphics[width=\textwidth]{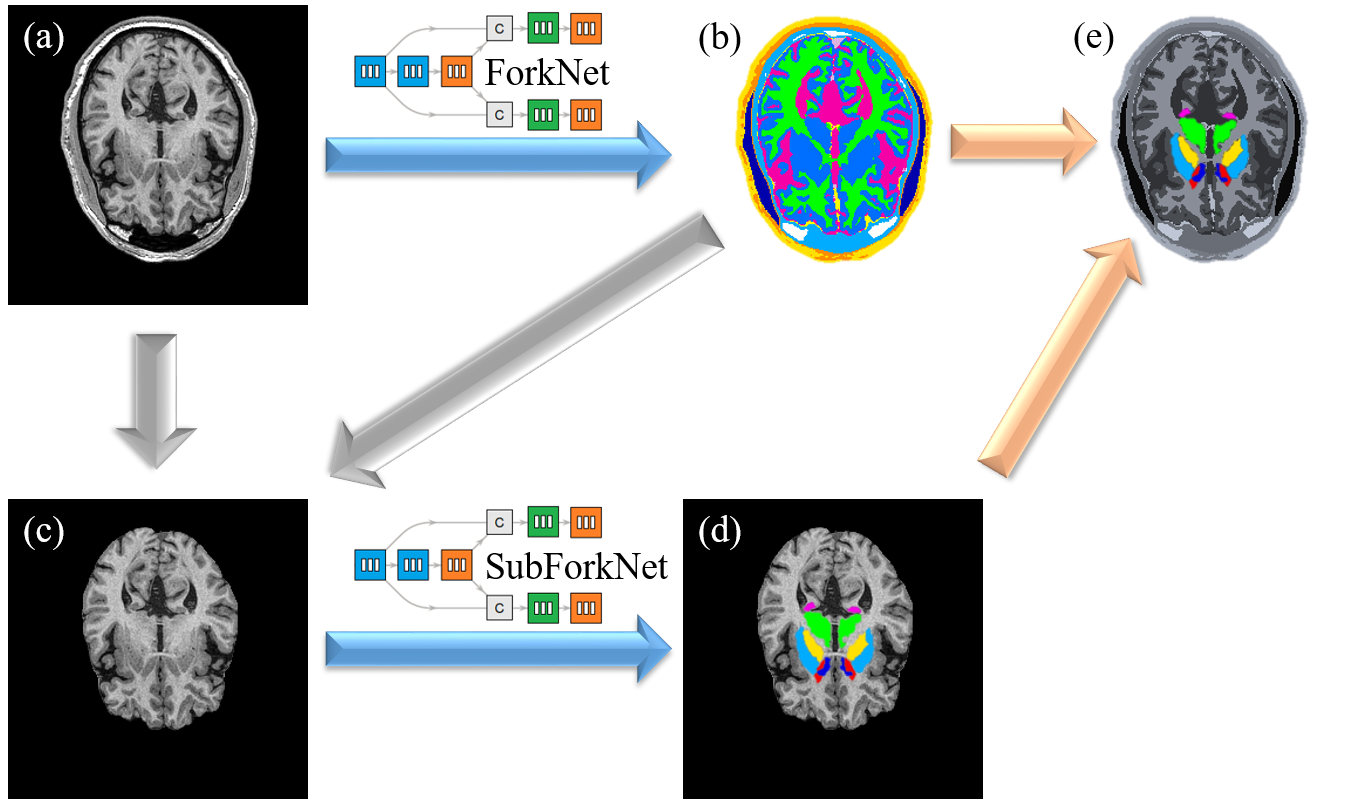}
\caption{Overview of the personalized head model pipeline. Structural MRI in (a) is segmented into different head tissues in (b) using ForkNet \citep{Rashed2019Neuroimage}. The segmented model is used for brain extraction in (c). Brain MRI is input to the SubForkNet for deep brain segmentation in (d). Finally, (d) is embedded in (b) to generate the final personalized head model in (e).} 
\label{pipeline}
\end{figure*}

To simulate the EF distribution in brain stimulation procedures with acceptable accuracy, a whole head model considering major tissues is required. This is a difficult task as several head tissues need to be segmented to be associated with equivalent tissue conductivity value. In our previous work, we proposed a deep learning method for segmentation of 13 head tissues \citep{Rashed2019Neuroimage} for cortical stimulation. However, that study did not consider the deep brain regions where the segmentation is more challenging and become essential in deep brain regions targeted by tDCS. Therefore, we present a pipeline for human head segmentation as follows. First, the 13 head tissues are annotated using ForkNet. Based on this segmentation, all non-brain tissues are excluded from the MRI (i.e., skull-stripping). Deep brain structures are then labeled using SubForkNet. This pipeline is demonstrated in Fig.~\ref{pipeline}.

\subsection{tDCS studies}

The electric potential generated by the current injected by the electrodes attached to the scalp was computed using the scalar potential finite-difference (SFPD) method with successive-over-relaxation and multigrid methods \citep{Dawson1998TMag, Laakso2012PMB}. First, the SPFD was used to solve the scalar-potential equation:

\begin{equation}
\nabla (\sigma \nabla \phi)=0,
\end{equation}
where $\phi$ and $\sigma$ denote the scalar potential and tissue conductivity, respectively. Then, the EF was obtained by dividing the potential between the two nodes along the edge of a cubic voxel (the minimum component of the model) by the length of the voxel edge. 

\begin{table*}
\centering
\footnotesize
\caption{Electric conductivity values of the head model tissues [S/m].}
\label{Conduct}
\setlength{\tabcolsep}{3pt}
\begin{tabular}{ |l c | l c |}
\hline
{\bf Tissue}	& {\bf Conductivity} &   {\bf Tissue}	& {\bf Conductivity}	  \\
\hline
\hline
Amygdala & 0.20 & Intervertebral disk & 0.10\\
Blood & 0.70  & Muscle &0.16\\
Bone (Cancellous) & 0.027 & Nucleus accumbens & 0.20\\
Bone (Cortical) & 0.008 &  Pallidum&0.20\\
Caudate & 0.20   & Putamen & 0.20 \\
Cerebellum &0.20 & Skin & 0.10 \\
CSF  & 1.80 & Thalamus&0.20\\
Fat &0.08 &Vitreous humor & 1.50\\
GM & 0.20 &  WM  & 0.14\\
Hippocampus & 0.20 & & \\
\hline
\end{tabular}
\end{table*}

\subsubsection{Electrode montages}
The tDCS electrode model was a 1-mm-thick rubber sheet (conductivity of 0.1 S/m) \citep{Saturnino2015neuroimage, Laakso2016neuroimage} inserted into a sponge soaked in normal saline solution (1.6 S/m) \citep{Dundas2007CN, Saturnino2015neuroimage}. The electrode was 5$\times$5 cm$^2$ and 5 mm of thickness. The injected current was 2 mA on top of the center of the rubber. Each electrode montage was a bipolar electrode (anode: positive pole and cathode: negative pole) placed at C3-Fp2 (motor cortex-supraorbital) positions according to the 10-20 electroencephalogram system. Large bipolar electrodes were selected to generate a large current spread in the brain, which is suitable for inducing high EFs in deep brain regions. Another common electrode size in clinics and research is 5$\times$7~cm$^2$. Though not used in this paper, we confirmed that the spatial distribution of the EF was marginally affected by the difference in the electrode size.

\subsubsection{Tissue conductivity}
The electrical conductivity of head tissues was assumed to be linear and isotropic, as shown in Table \ref{Conduct}, on the basis of the values reported in \cite{GomezTames2016PMB, Laakso2016neuroimage}.

\section{Results}

\subsection{Analysis methods}

The segmentation quality is evaluated using Dice Coefficient (DC) defined as follows:

\begin{equation}
DC(R,R_{\circ})=\frac{2|R \cap R_{\circ}|}{|R|+|R_{\circ}|} \times 100\%,
\end{equation}
where $R$ and $R_{\circ}$ are the network segmented volume and golden truth one, respectively. Moreover, the Hausdorff distance (HD) is used as a quality metric and is defined as:

\begin{equation}
HD(R,R_{\circ})=\max_{a \in R} [ \min_{b \in R_{\circ}} [ d(a,b]],
\end{equation}
where $d(.,.)$ is the Euclidian distance between the two voxels $a$ and $b$. To quantify the global difference of the internal EF distributions obtained by the different segmented head models and the ground truth, the normalized average of point-wise absolute difference (global error) is used as follows:

\begin{equation}
\textnormal{Diff}(R,R_{\circ})=\frac{1}{\max_{i \in \Omega} (E(i), E_{\circ}(i))}\times \frac{\sum_{i=1}^I  |E(i)-E_{\circ}(i)|}{I} \times 100\%,
\end{equation}
where $E$ and $E_\circ$ are the internal EF in $R$ and $R_\circ$ models, respectively. The relative difference of the maximum internal EF is used as of the local error. These metrics are applied to the different deep brain structures separately and also to the brain (white matter and grey matter) and the whole deep brain tissues. To mitigate numerical artifacts derived from computing the EF using the voxel model at the surface of the CSF-brain boundaries \citep{Reilly2016PMB}, post-processing based on the 99.9$^{th}$ percentile value of the EF was applied for each tissue \citep{GomezTames2018TEMC}. In this work, the EF strength was adopted as metric of neuromodulation to demonstrate the accuracy of the proposed segmentation. Although the most suitable metric is still to be resolved \citep{Laakso2019SR, Antonenko2019BS}.

\begin{figure*}
\centering
\includegraphics[width=\textwidth]{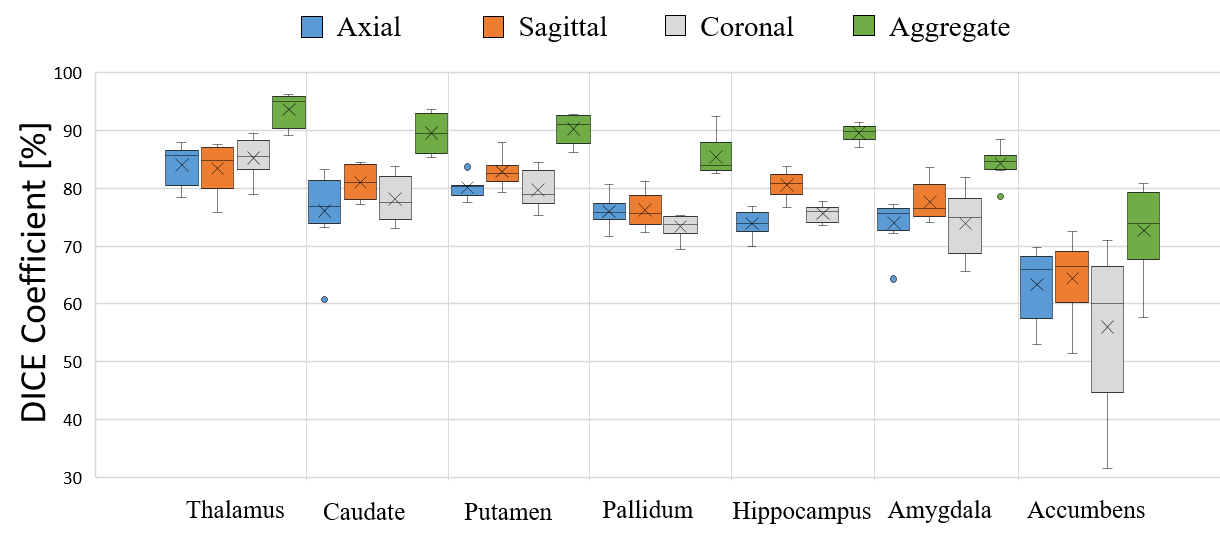}
\caption{Boxplot of DC values computed from 8 subjects segmented using SubForkNet. Cross indicates the mean value, center line indicates the median, box indicates the first and third quartiles, whiskers indicate the maximum and minimum values, and circles are the outliers. A sample of the segmentation results corresponding to aggregate model is shown in Fig.~\ref{s1}}.
\label{aggregate}
\end{figure*}

\begin{figure*}
\centering
\includegraphics[width=\textwidth]{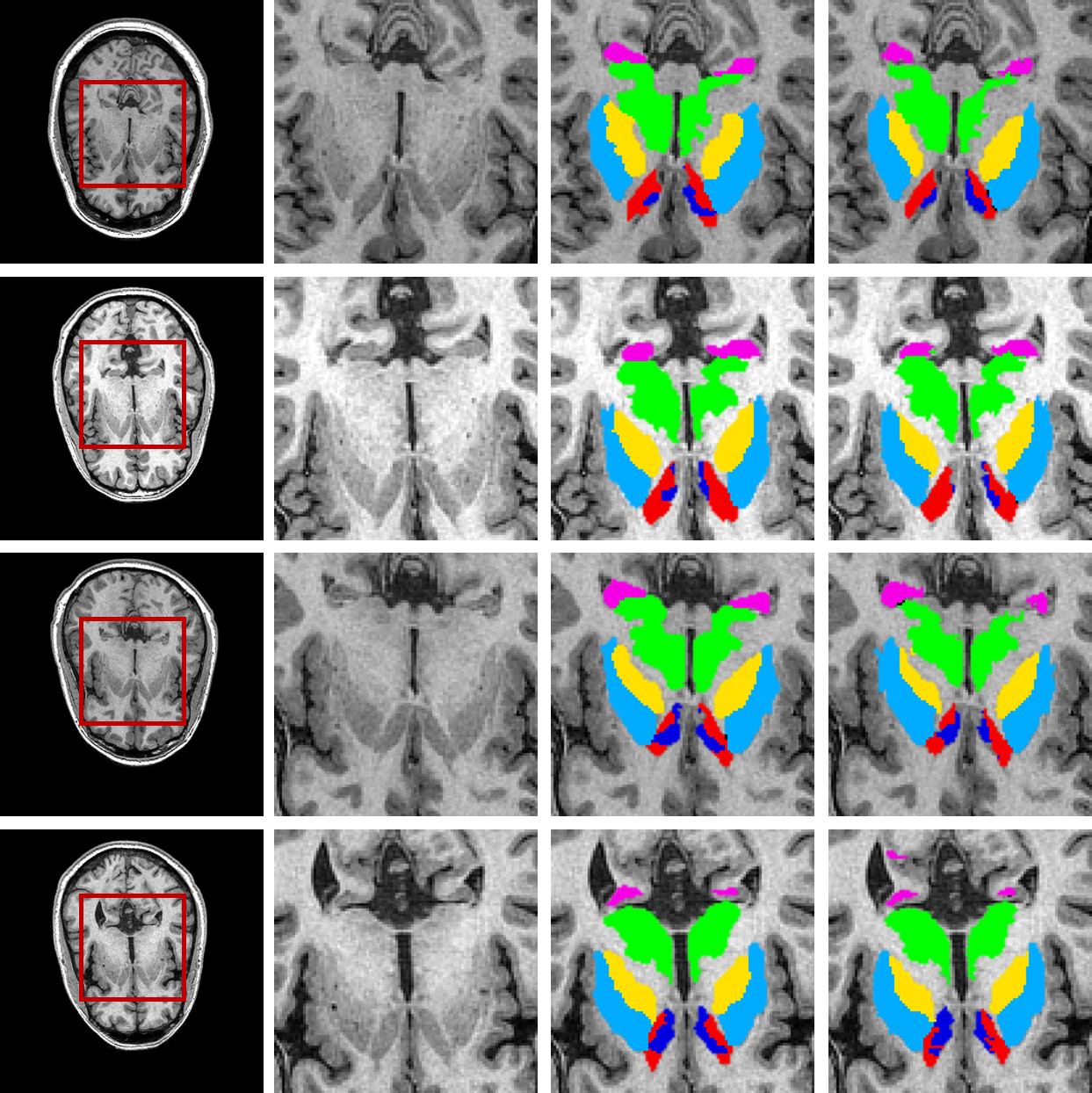}
\caption{From left to right, T1-weighted MRI axial slice, magnified ROI region, golden truth segmentation ($R_{\circ}$), and SubForkNet segmentation. From top to bottom, subjects case01017, case01019, case01025, and case01028. Color code of deep regions is the same as in Fig.~\ref{example}.} 
\label{s1}
\end{figure*}

\subsection{Deep brain segmentation (NAMIC dataset)}

The NAMIC dataset is randomly divided into 10 subjects for training and 8 subjects for testing. The training subjects are used to extract slices in a different directions, and then the whole set of slices is shuffled and split into 90\% for training and 10\% for validation. The proposed architecture is implemented using Wolfram Mathematica (R) ver. 12.0, installed on a workstation of 4$\times$Intel (R) Xeon CPUs @ 3.60 GHz, 128 GB memory, and 3$\times$NVIDIA GeForce GTX 1080 GPUs. In training, the cross-entropy loss function is minimized using ADAM algorithm \citep{Kingma2014arXiv}. Network initialization and learning rate are computed automatically by the \texttt{NetInitialize} and \texttt{NetTrain} Mathematica functions, respectively. The network training is conducted using 100 epochs with batch size = 4. Three networks are used in different directions with unified degree $N=7$, depth $D=2$, $\epsilon=0.3$, and convolutional window $r^2=3\times3$. Segmentation results are evaluated using the semi-automatic segmentation as a golden truth, and DC values are shown in Fig.~\ref{aggregate}. A sample of the segmented deep brain structures is shown in Fig.~\ref{s1}. It is observed that segmentation of Thalamus, Caudate, and Putamen structures is of high quality, which is relatively recognized from being presented in a relatively large region in MRI. However, small regions such as Accumbens structure are of low segmentation quality. Another observation is the superior quality of aggregate segmentation, which considers the combination of segmentation along the three directions and elimination of voxels located outside GM regions.

\begin{figure*}
\centering
\includegraphics[width=\textwidth]{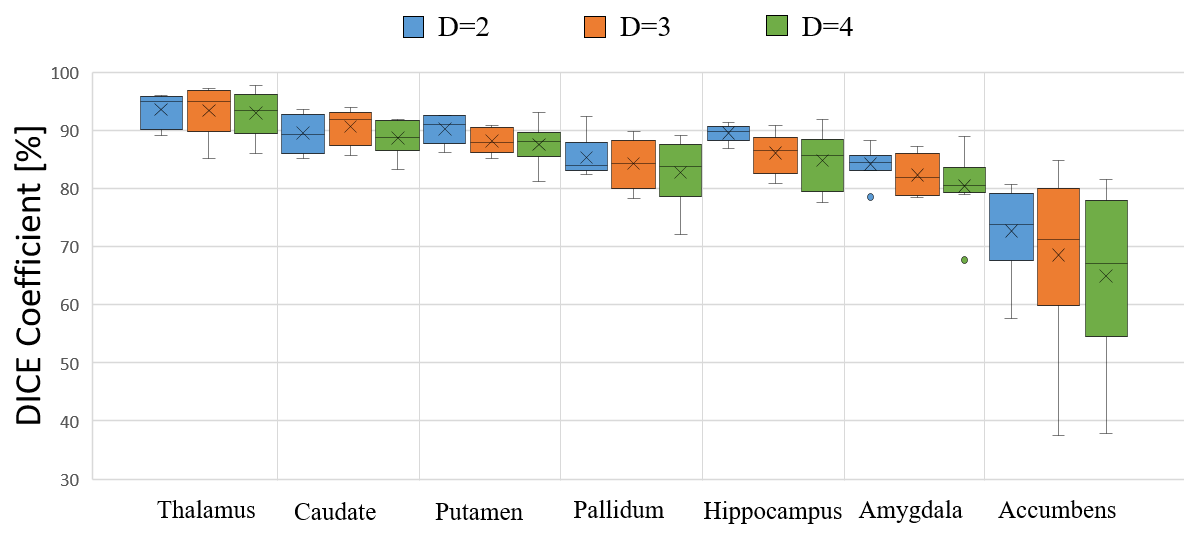}
\caption{Boxplot of DC values computed from 8 subjects using SubForkNet with different depth $D$ values.} %
\label{depth}
\end{figure*}

\begin{figure*}
\centering
\includegraphics[width=\textwidth]{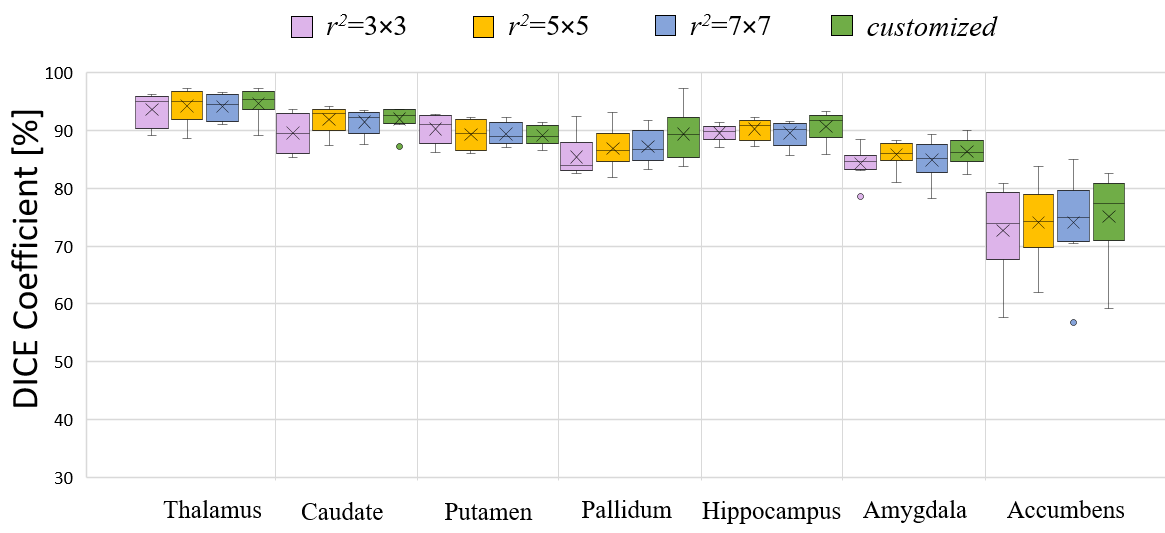}
\caption{Boxplot of DC values computed from 8 subjects using SubForkNet with different convolution kernel size $r^2$. Details of customized kernel size are listed in Table~\ref{Dicecomp2}.} %
\label{kernel}
\end{figure*}

Other studies were conducted to understand how different architecture parameters are related to the segmentation quality. Similar to previous study, network is trained over 100 epochs with batch size of 4. First, we evaluate segmentation obtained using different network depth ($D=2$, $3$, and $4$). Although more deep network means more observed features, but we have found that going deeper may not have a good influence on the DC coefficient as shown in Fig.~\ref{depth} and Table~\ref{Dicecomp1}. One reason might be the over-fitting of the trained network. Also, another study is used to evaluate different kernel size ($r^2$) of convolution operation. Segmentation is conducted using architectures of 3$\times$3, 5$\times$5, and 7$\times$7 kernel size. Moreover, a new architecture is customized based on the observation of results obtained from different kernel size on different slicing directions. Superior DC values are used to design a customized kernel size for each decoder independently as detailed in Table~\ref{Dicecomp2}. The measured dice values are shown in Fig.~\ref{kernel} and it indicates that the use of customized design can slightly improve the segmentation quality.    


\begin{table*}
\centering
\footnotesize
\caption{Mean and standard deviation of Dice coefficient values for deep barin structures computed from 8 subjects with different network depth setups and fixed kernel size ($r^2=3 \times 3$). Bold indicate superior mean value.}
\label{Dicecomp1}
\setlength{\tabcolsep}{3pt}
\begin{tabular}{| l| c  c  c |}
\hline

 	{\bf Structure} & $D=2$ &  $D=3$ & $D=4$ \\

\hline
\hline
Thalamus   & {\bf 93.65} $\pm$ 2.86 &  93.52 $\pm$ 4.35 & 93.08 $\pm$ 3.99 	\\
Caudate    & 89.59 $\pm$ 3.28 & {\bf 90.75} $\pm$ 3.10 & 88.79 $\pm$ 3.03 \\
Putamen    & {\bf 90.31} $\pm$ 2.52 & 88.22 $\pm$ 2.32 & 87.63 $\pm$ 3.55 \\
Pallidum   & {\bf 85.41} $\pm$ 3.48 & 84.73 $\pm$ 4.16 & 82.80 $\pm$ 5.96 \\
Hippocampus& {\bf 89.62} $\pm$ 1.45 & 86.20 $\pm$ 3.48 & 84.90 $\pm$ 5.00 \\
Amygdala   & {\bf 84.27} $\pm$ 2.82 & 82.43 $\pm$ 3.72 & 80.51 $\pm$ 6.03 \\
Accumbens  & {\bf 72.71} $\pm$ 7.79 & 68.63 $\pm$ 15.19&	64.98 $\pm$14.56 \\
\hline \hline
Deep regions & {\bf 86.51} $\pm$ 7.39 & 84.87 $\pm$ 9.81 & 83.24 $\pm$ 10.75 \\
\hline 
\end{tabular}
\end{table*}


\begin{table*}
\centering
\footnotesize
\caption{Mean and standard deviation of Dice coefficient values for deep brain structures computed from 8 subjects of different network kernel sizes and fixed depth ($D=2$). Bold indicate superior mean value.}
\label{Dicecomp2}
\setlength{\tabcolsep}{3pt}
\begin{tabular}{| l| c | c | c |c c c c|}
\hline
 \multirow{2}{*}{\bf Structure} 	& \multicolumn{3}{|c|}{\bf Customized kernel} &\multirow{2}{*}{$r^2=3 \times 3$}& \multirow{2}{*}{$r^2=5 \times 5$}&\multirow{2}{*}{$r^2=7 \times 7$}& \multirow{2}{*}{Customized}\\
 \cline{2-4}
  	& axial &  sagittal & coronal &  & &  & \\

\hline
\hline
Thalamus   & $5 \times 5$ & $7 \times 7$ & $7 \times 7$ & 93.65 $\pm$ 2.86 & 94.28 $\pm$ 2.99 & 94.13 $\pm$ 2.22 & {\bf 94.70} $\pm$ 2.57	\\
Caudate    &$5 \times 5$& $5 \times 5$ & $5 \times 5$ &  89.59 $\pm$ 3.28 & 91.93 $\pm$ 2.32 & 91.47 $\pm$ 2.16 & {\bf 91.98} $\pm$ 2.16\\
Putamen    & $3 \times 3$& $5 \times 5$ & $5 \times 5$ &  {\bf 90.31} $\pm$ 2.52 & 89.29 $\pm$ 2.54 & 89.37 $\pm$ 1.95 & 89.15 $\pm$ 1.72\\
Pallidum   & $7 \times 7$ & $7 \times 7$ & $7 \times 7$ &  85.41 $\pm$ 3.48 & 86.88 $\pm$ 3.46	&  87.20 $\pm$ 2.93 & {\bf 89.42} $\pm$ 4.41\\
Hippocampus& $7 \times 7$ & $5 \times 5$ & $5 \times 5$ &  89.62 $\pm$ 1.45 & 90.29 $\pm$ 1.93 & 89.55 $\pm$ 2.26 & {\bf 90.74} $\pm$ 2.56\\
Amygdala   & $7 \times 7$ & $5 \times 5$ & $5 \times 5$ &  84.27 $\pm$ 2.82 &  85.83 $\pm$ 2.34 & 84.94 $\pm$ 3.46 & {\bf 86.34} $\pm$ 2.39\\
Accumbens  & $5 \times 5$ & $7 \times 7$ & $5 \times 5$ &  72.71 $\pm$ 7.79 & 74.05 $\pm$ 6.81 &  74.13 $\pm$ 8.49 & {\bf 75.13} $\pm$ 7.62	\\
\hline \hline
\multicolumn{4}{|l|}{Deep regions} &  86.51 $\pm$ 7.39 & 87.51 $\pm$ 7.01 & 87.26 $\pm$ 7.13 & {\bf 88.21} $\pm$ 6.93\\
\hline 
\end{tabular}
\end{table*}

\begin{figure*}
\centering
\includegraphics[width=\textwidth]{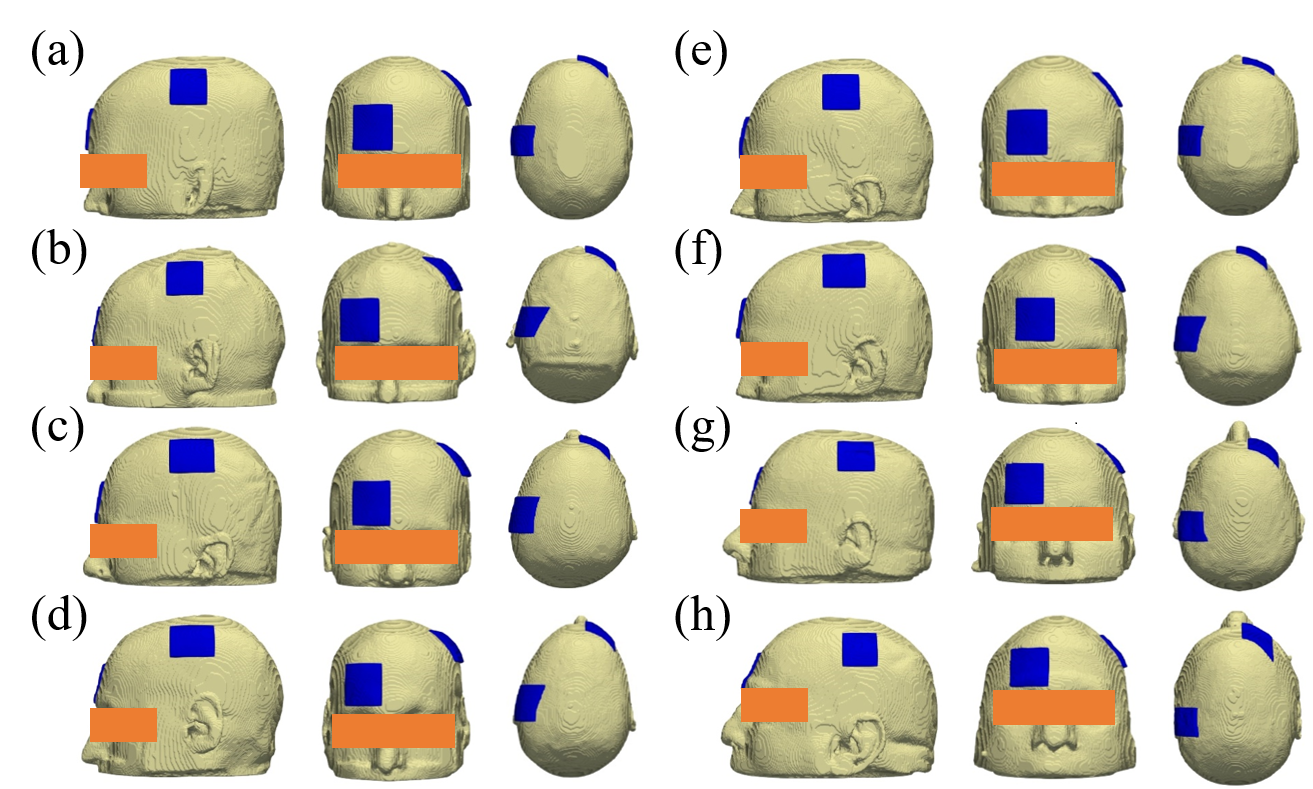}
\caption{tDCS C3-Fp2 montage for subjects (a) case01017, (b) case01019, (c) case01025, (d) case01028, (e) case01034, (f) case01039, (g) case01042, and (h) case01045. All subjects are shown in lateral, frontal and top views, from left to right. Blue markers indicate electrode position.} %
\label{montage}
\end{figure*}

\subsection{Deep brain segmentation (MICCAI 2012 dataset)}
For further validation of the proposed method, the well-known MICCAI 2012 dataset is used for additional evaluation. This dataset is divided into 15 subjects for training and 20 subjects for testing. The segmentation pipeline descriped in Fig.~\ref{pipeline} is used and the labels corresponding to left and right portions of each deep brain structures are unified. SubForkNet with $D=2$, $N=7$, $r^2=3\times 3$, $\epsilon=0.3$ is trained using 100 epochs and batch size=4. Dice coefficient and HD values are shown in Tables~\ref{Dicecomp3} and~\ref{HDcomp3} with corresponding values computed using FIRST~\citep{Patenaude2011neuroimage}, FreeSurfer~\citep{Fischi2012NeuroImage}, PICSL method \citep{Wang2013MICCAI} and CNN+Atlas method \citep{Kushibar2018MIA}. Although the network parameters are not optimized to achieve the best performance, it still can provide notable improvement compared with related methods. 


\begin{table*}
\centering
\footnotesize
\caption{Mean and standard deviation of Dice coefficient values for deep brain structures computed from 20 subjects (MICCAI 2012 dataset) using FIRST~\citep{Patenaude2011neuroimage}, FreeSurfer~\citep{Fischi2012NeuroImage}, PICSL \citep{Wang2013MICCAI}, CNN+Atlas \citep{Kushibar2018MIA}, and the proposed method. Results of previous methods demonstrate segmentation of left (L) and right (R) structures independently.}
\label{Dicecomp3}
\setlength{\tabcolsep}{3pt}
\begin{tabular}{| l c| ccccc |}
\hline
{\bf Structure} 	& & FIRST & FreeSurfer &PICSL  & CNN+Atlas & Proposed \\

\hline
\hline
\multirow{2}{*}{Thalamus} & L  & 88.9 $\pm$ 1.8 &83.0 $\pm$ 1.8 &92.2 $\pm$ 1.3 & 92.1 $\pm$ 1.8 & \multirow{2}{*}{{\bf 92.6} $\pm$ 9.0} 	\\
 & R  &  89.0 $\pm$ 1.7 &84.9 $\pm$ 2.1& 92.4 $\pm$ 0.8 & 92.0 $\pm$ 1.6 & 	\\
\hline
\multirow{2}{*}{Caudate}    & L &  79.7 $\pm$ 4.6 &80.8 $\pm$ 7.9 & 88.5 $\pm$ 7.4 & {\bf 89.4} $\pm$ 7.1 & \multirow{2}{*}{87.4 $\pm$ 19.6}\\
 & R  & 83.7 $\pm$ 11.7 & 80.1 $\pm$ 4.2 & 88.7 $\pm$ 6.5 & {\bf 89.2} $\pm$ 5.7 & 	\\
 \hline
\multirow{2}{*}{Putamen}    & L & 86.0 $\pm$ 6.0 & 77.1 $\pm$ 3.9 & 90.9 $\pm$ 4.2 & 91.6 $\pm$ 2.3 & \multirow{2}{*}{{\bf 93.5} $\pm$ 6.5}\\
 & R  &   87.6 $\pm$ 8.0 & 79.9 $\pm$ 2.6 & 90.8 $\pm$ 4.6 & 91.4 $\pm$ 3.1 & 	\\
\hline
\multirow{2}{*}{Pallidum}   & L &  81.5 $\pm$ 8.8 & 69.3 $\pm$ 18.9 & 87.3 $\pm$ 3.2 & 84.3 $\pm$ 10.1 & \multirow{2}{*}{{\bf 90.0} $\pm$ 7.1}\\
 & R  &  79.9 $\pm$ 6.0 & 79.2 $\pm$ 8.5 & 87.4 $\pm$ 4.7 & 86.1 $\pm$ 4.9 & 	\\
 \hline
\multirow{2}{*}{Hippocampus} & L & 80.9 $\pm$ 2.2 & 78.4 $\pm$ 5.4 & 87.1 $\pm$ 2.4 & 87.6 $\pm$ 2.0 & \multirow{2}{*}{{\bf 88.1} $\pm$ 6.6}\\
 & R  &  81.0 $\pm$ 14.0 & 79.4 $\pm$ 2.5 & 86.9 $\pm$ 2.2 & 87.9 $\pm$ 2.0 & 	\\
 \hline
\multirow{2}{*}{Amygdala}   & L &   72.1 $\pm$ 5.3 & 58.5 $\pm$ 6.4 & 83.2 $\pm$ 2.6 & 83.3 $\pm$ 3.2 & \multirow{2}{*}{{\bf 85.4} $\pm$ 5.6}\\
 & R  &   70.7 $\pm$ 5.4 & 57.6 $\pm$ 7.6 & 81.2 $\pm$ 3.3 & 82.1 $\pm$ 2.7 & 	\\
 \hline
\multirow{2}{*}{Accumbens}  & L &  69.9 $\pm$ 8.9 & 63.0 $\pm$ 5.5 & 79.0 $\pm$ 5.0 & {\bf 79.9} $\pm$ 5.2 & \multirow{2}{*}{78.6 $\pm$ 15.2}\\
 & R  &  67.8 $\pm$ 8.1 & 44.3 $\pm$ 6.5 & 78.3 $\pm$ 5.8 & {\bf 79.1} $\pm$ 6.7 & 	\\
 
\hline \hline
Deep regions &  & 79.9 $\pm$ 9.4 & 72.5 $\pm$ 13.7 &86.7 $\pm$ 6.1 & 86.9 $\pm$ 6.4 & {\bf 87.9} $\pm$ 11.8\\
\hline 
\end{tabular}
\end{table*}


\begin{table*}
\centering
\footnotesize
\caption{Mean and standard deviation of HD values for deep brain structures computed from 20 subjects (MICCAI 2012 dataset) using FIRST, FreeSurfer, PICSL, CNN+Atlas, and the proposed method. Results of previous methods demonstrate segmentation of left (L) and right (R) structures independently.}
\label{HDcomp3}
\setlength{\tabcolsep}{3pt}
\begin{tabular}{| l c| ccccc |}
\hline
{\bf Structure} 	& & FIRST & FreeSurfer &PICSL  & CNN+Atlas & Proposed \\

\hline
\hline
\multirow{2}{*}{Thalamus} & L & 4.65 $\pm$ 0.90 & 4.94 $\pm$ 1.01 & 3.22 $\pm$ 0.99 & 3.39 $\pm$ 1.13 & \multirow{2}{*}{{\bf 2.49} $\pm$ 2.20} 	\\
                          & R & 4.39 $\pm$ 0.92 & 4.76 $\pm$ 0.75 & 3.11 $\pm$ 0.79 & 3.31 $\pm$ 1.01 & 	\\
\hline
\multirow{2}{*}{Caudate}  & L & 3.56 $\pm$ 1.30 & 9.89 $\pm$ 3.09 & 3.44 $\pm$ 1.89 & {\bf 3.32} $\pm$ 2.00 & \multirow{2}{*}{3.95 $\pm$ 4.65}\\
                          & R & 4.16 $\pm$ 1.37 &10.39 $\pm$ 3.09 & 3.60 $\pm$ 1.67 & {\bf 3.51} $\pm$ 1.67 & \\
\hline
\multirow{2}{*}{Putamen}  & L & 3.79 $\pm$ 1.76 & 6.31 $\pm$ 1.09 & 3.07 $\pm$ 1.40 & {\bf 2.63} $\pm$ 1.09 & \multirow{2}{*}{ 3.22 $\pm$ 2.09}\\
                          & R & 3.26 $\pm$ 1.23 & 5.85 $\pm$ 0.84 & 2.91 $\pm$ 1.41 & {\bf 2.75} $\pm$ 0.99 & \\
\hline
\multirow{2}{*}{Pallidum} & L & 2.89 $\pm$ 0.71 & 3.89 $\pm$ 1.07 & 2.52 $\pm$ 0.54 & 2.38 $\pm$ 0.76 & \multirow{2}{*}{{\bf 1.90} $\pm$ 0.60}\\
                          & R & 3.18 $\pm$ 0.93 & 3.45 $\pm$ 0.98 & 2.49 $\pm$ 0.59 & 2.59 $\pm$ 0.61 & 	\\
\hline
\multirow{2}{*}{Hippocampus} & L & 5.49 $\pm$ 1.66 &6.35 $\pm$ 1.87 & 4.34 $\pm$ 1.66 & 4.48 $\pm$ 2.02 & \multirow{2}{*}{{\bf 2.92} $\pm$ 0.75}\\
                          & R & 4.80 $\pm$ 1.66 & 6.19 $\pm$ 1.59 & 4.01 $\pm$ 1.45 & 3.76 $\pm$ 1.23 & 	\\
\hline
\multirow{2}{*}{Amygdala} & L &  3.54 $\pm$ 0.72 & 5.05 $\pm$ 0.97 & {\bf 2.44} $\pm$ 0.29 & 2.39 $\pm$ 0.39 & \multirow{2}{*}{ 2.88 $\pm$ 2.39}\\
                          & R & 4.11 $\pm$ 0.75 & 5.43 $\pm$ 0.90 & {\bf 2.72} $\pm$ 0.50 & 2.72 $\pm$ 0.69 & 	\\
\hline
\multirow{2}{*}{Accumbens}& L & 6.81 $\pm$ 8.76 & 4.28 $\pm$ 1.11 & 2.57 $\pm$ 0.67 & 2.39 $\pm$ 0.64 & \multirow{2}{*}{{\bf 2.33} $\pm$ 0.87}\\
                          & R & 3.93 $\pm$ 1.75 & 5.47 $\pm$ 1.02 & 2.65 $\pm$ 0.76 & 2.54 $\pm$ 0.65 & 	\\
\hline \hline
Deep regions &  & 4.18 $\pm$ 2.76 & 5.87 $\pm$ 2.48 & 3.08 $\pm$ 1.27 & 3.01 $\pm$ 1.30 & {\bf 2.83} $\pm$ 2.36\\
\hline 
\end{tabular}
\end{table*}
\subsection{tDCS validation}

\begin{figure*}
\centering
\includegraphics[width=\textwidth]{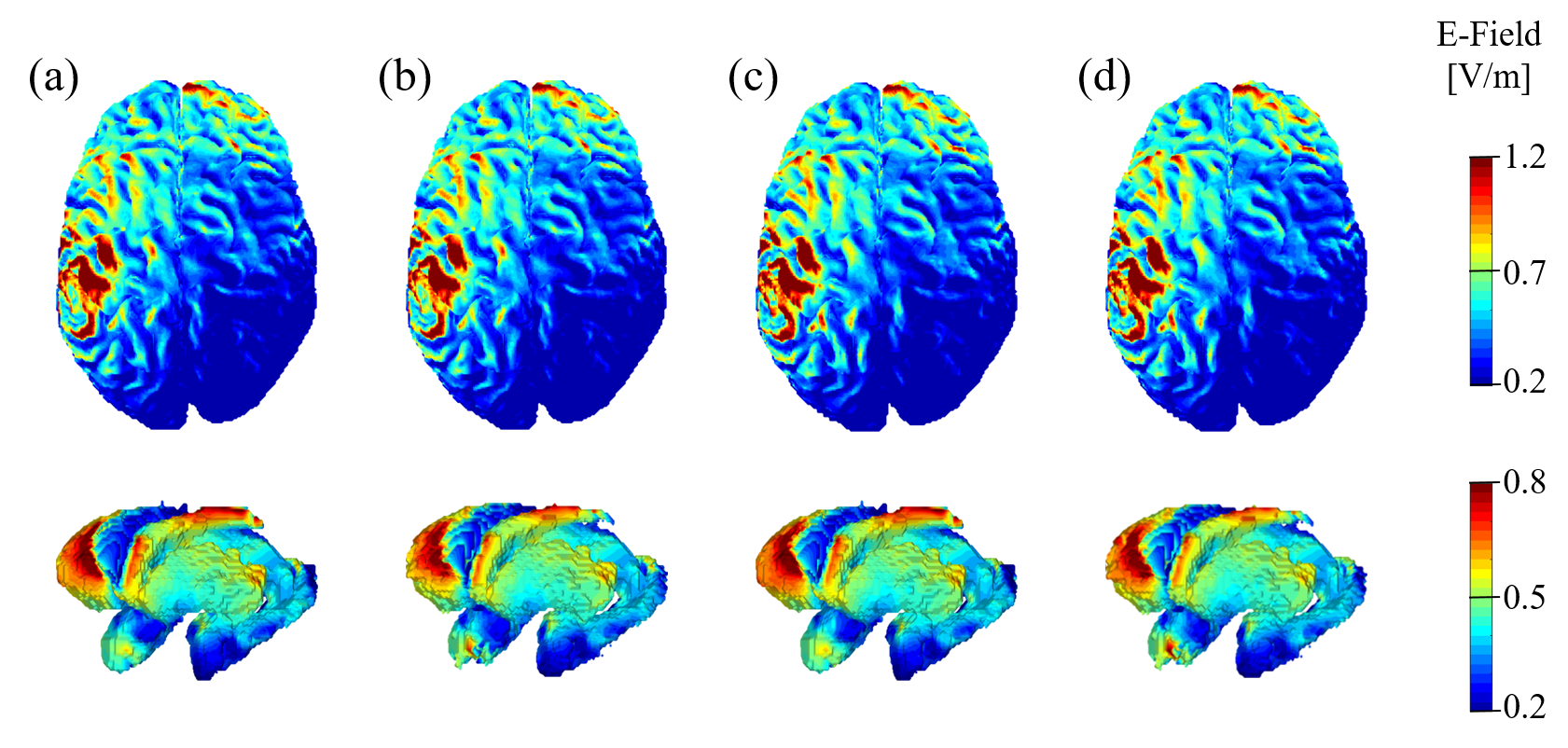}
\caption{Electric field distribution (EF) in brain (top) and deep brain regions (bottom) for models (a) $R_s^s$, (b) $R_s^f$, (c) $R_f^s$, and (d) $R_f^f$ of subject case01017.} 
\label{tDCS}
\end{figure*}

To validate the effect of the deep brain structure's segmentation using SubForkNet and how it is related with the tDCS measurements, we generated four different head models; $R_s^s$, $R_s^f$, $R_f^s$, and $R_f^f$. Where $R_a^b$ is the head model generated from 13 head tissues segmentation using method $a$ and deep brain structure's segmentation using method $b$, $s$ refers to the semi-automatic method (golden truth) and $f$ refers to ForkNet or SubForkNet segmentation. The tDCS electrodes positioning for eight subjects are shown in Fig.~\ref{montage}. The four head models generated for the subject (case01017, NAMIC dataset) were used to carry out tDCS studies. The EF induced in the brain and deep brain structures is shown in Fig.~\ref{tDCS}. From this figure, even where some segmentation errors can be observed in deep brain structures, their impact on the tDCS simulation results was insignificant. Hot spot regions in cortical surface and deep regions are of high consistency. The tDCS simulation is repeated for the remaining seven subjects to investigate the variability factors and quantitative evaluation results are shown in Table~\ref{tDCSStat}.  Global error can reach mean of 3.11\% in Amygdala structure in case both 13 head tissues and deep regions are segmented using ForkNet and SubForkNet, respectively (i.e.~$R_f^f$). As for the local error, Accumbens, Hippocampus, and Amygdala record the mean values of 12.91\%, 11.00\%, and 7.79\%, respectively. These values are relatively small considering that it represent the segmentation error in both major head tissues and deep brain structures. Smaller values are observed when the segmentation of main head tissues are set to the golden truth values (i.e.~$R_s^f$).

\begin{table*}
\centering
\footnotesize
\caption{Global and local electric field error values computed from 8 subjects for different brain regions and different head models compared to golden truth $R_s^s$.}
\begin{tabular}{|l|ccc|ccc|}
\hline 
\multirow{2}{*}{\bf Region} &\multicolumn{3}{c|}{\bf Global Error}  & \multicolumn{3}{c|}{\bf Local Error}\\
 \cline{2-7}
 & $R_s^f$ & $R_f^s$ & $R_f^f$& $R_s^f$ & $R_f^s$ & $R_f^f$ \\
\hline \hline
Thalamus    & 0.04  $\pm$ 0.11  & 2.10  $\pm$ 1.17  & 2.31  $\pm$ 1.19 
            & 4.15  $\pm$ 5.64  & 4.37  $\pm$ 4.41  & 6.30  $\pm$ 4.86 \\
Caudate     & 0.03  $\pm$ 0.08  & 1.84  $\pm$ 1.13  & 2.16  $\pm$ 1.17
            & 10.54 $\pm$ 7.94  & 4.32  $\pm$ 5.11  & 9.56  $\pm$ 6.30 \\
Putamen     & 0.06  $\pm$ 0.16  & 1.75  $\pm$ 1.57  & 1.88  $\pm$ 1.54
            & 2.38  $\pm$ 1.37  & 2.78  $\pm$ 3.35  & 2.94  $\pm$ 2.96 \\
Pallidum    & 0.07  $\pm$ 0.20  & 1.71  $\pm$ 1.52  & 1.80  $\pm$ 1.45
            & 2.88  $\pm$ 3.26  & 2.77  $\pm$ 2.10  & 5.60  $\pm$ 3.77 \\
Hippocampus & 0.08  $\pm$ 0.21  & 2.76  $\pm$ 0.63  & 2.87  $\pm$ 0.80
            & 7.92  $\pm$ 11.35 & 4.31  $\pm$ 1.86  & 11.00 $\pm$ 12.25 \\
Amygdala    & 0.09  $\pm$ 0.24  & 3.10  $\pm$ 0.87  & 3.11  $\pm$ 0.92
            & 3.23  $\pm$ 4.80  & 3.71  $\pm$ 2.34  & 7.79  $\pm$ 7.90 \\
Accumbens   & 0.05  $\pm$ 0.13  & 1.74  $\pm$ 1.17  & 1.82  $\pm$ 1.11
            & 12.19 $\pm$ 8.42  & 4.35  $\pm$ 3.99  & 12.91 $\pm$ 8.22 \\
\hline \hline
Deep regions & 0.05 $\pm$ 0.90 & 1.69 $\pm$ 0.90 & 1.85 $\pm$ 0.90
             & 6.30 $\pm$ 4.66 & 2.87 $\pm$ 3.78 & 7.69 $\pm$ 4.79\\
Brain        & 0.00 $\pm$ 0.00 & 0.98 $\pm$ 0.41 & 0.98 $\pm$ 0.41
             & 0.00 $\pm$ 0.00 & 4.47 $\pm$ 3.87 & 4.47 $\pm$ 3.86\\
\hline 
\end{tabular}
\label{tDCSStat}
\end{table*}

\section{Discussion}

\subsection{SubForkNet architecture}

In our previous study \citep{Rashed2019Neuroimage}, we proposed the ForkNet architecture for automatic segmentation of major head tissues. Then, the EF distribution in the cortical region of the brain was computed. It was found that the segmentation accuracy of peripheral structures such as skin, scalp, and CSF were significantly important compared to deep regions for the accurate estimation of TMS-induced EFs in cortical regions. In this study, we investigated the segmentation of deep brain tissues using semantic end-to-end network architecture, and then studied how it is correlated to the accuracy of induced EF. In contrast to TMS, tDCS-generated EFs are significant not only on superficial but also on deeper brain regions and highly sensitive to the segmentation accuracy of deep structures and surrounding regions. Therefore it is important to investigate how the proposed network architecture can provide a segmentation quality that is reliable for personalized tDCS studies.

The network architecture presented here is proved to generate a good segmentation quality using only T1-weighted MRI. One important feature is the split design of decoders, which provide a more feasible architecture for network customization. In this study, we presented a single example of this feature, where a multi-size convolution kernel can achieve better segmentation accuracy, as shown in Table~\ref{Dicecomp2}. The personalized head model pipeline shown in Fig.~\ref{pipeline} demonstrate how the two networks ForkNet and SubForkNet (proposed) can be combined for rigid segmentation of different head tissues. Both networks are designed for different segmentation tasks. The former is for superficial and the latter is for deep regions. Customization of the proposed network architecture, such as network depth and convolutional kernel size, were evaluated. The tDCS neuromodulation in deep brain regions was computed for the generated head models and compared with those generated using alternative methods.

\begin{figure*}
\centering
\includegraphics[width=\textwidth]{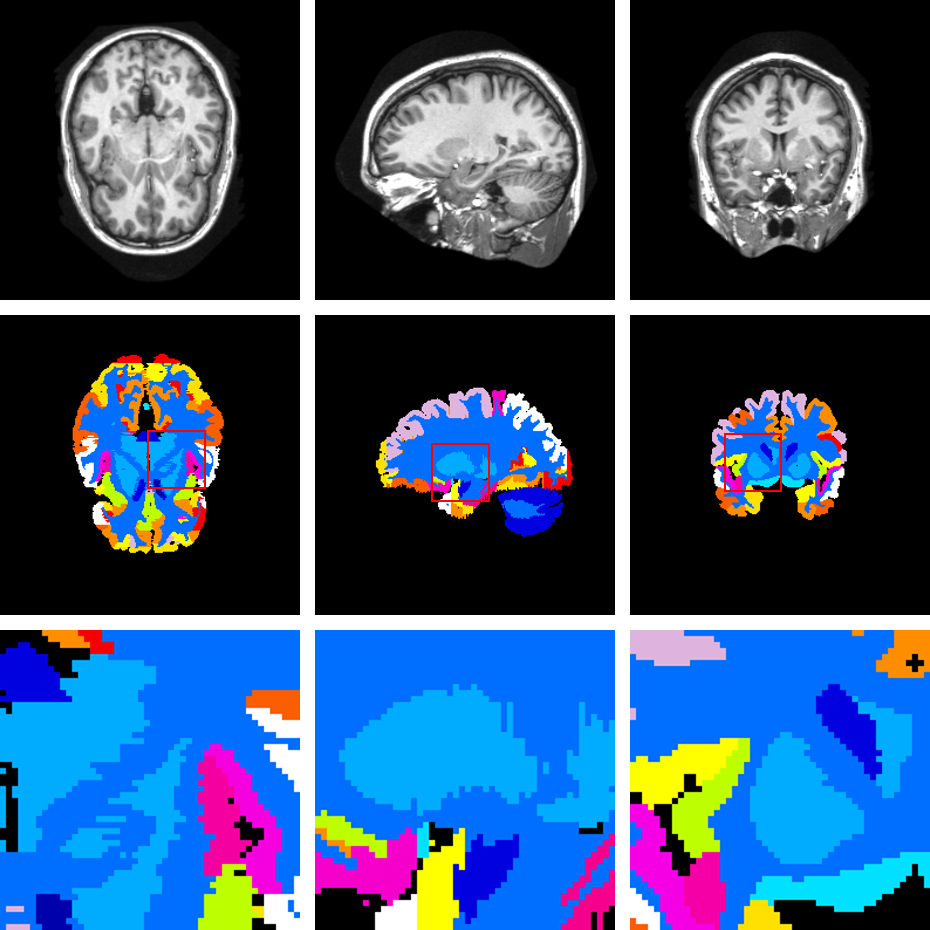}
\caption{Example demonstrate segmentation accuracy of MICCAI 2012 dataset (subject: 1003). From top to bottom, MRI, golden truth segmentation, and magnification of labeled region. Slices in axial, sagittal, and coronal are shown from left to right.} 
\label{datalimit0}
\end{figure*}

\begin{figure*}
\centering
\includegraphics[width=\textwidth]{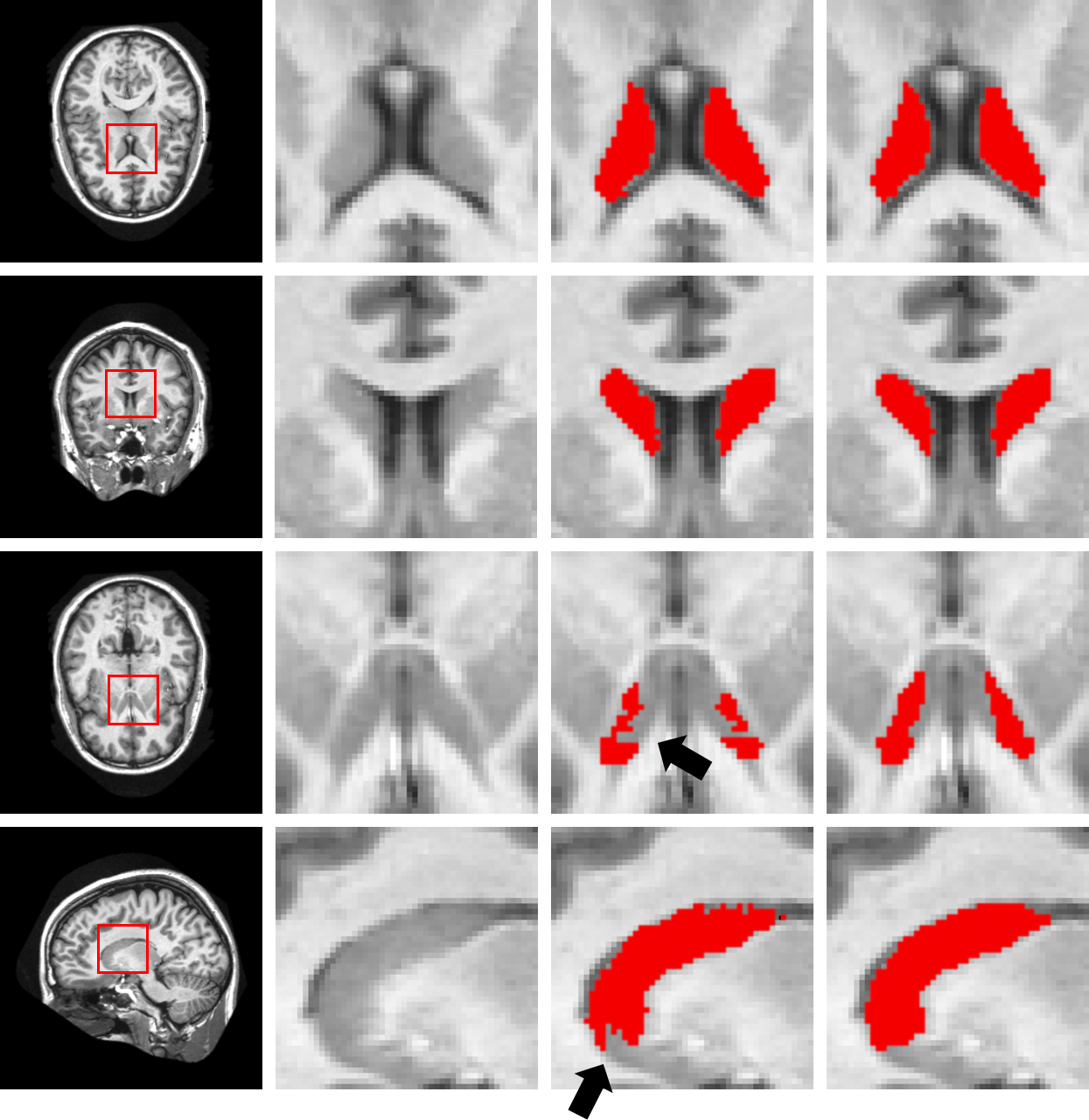}
\caption{Example demonstrate challenges related to the training data in MICCAI 2012 dataset (subject: 1003). From left to right, MRI slice, magnification of labeled region, golden truth segmentation and SubForkNet segmentation of Caudate structure. First two rows demonstrate examples where segmentation is highly matched. Latter ones, demonstrate examples where mismatch can be easily recognized (indicated by arrows), however, SubForkNet looks more accurate referring to anatomical image.} 
\label{datalimit}
\end{figure*}

\subsection{Data limitations}

The performance of supervised-based segmentation is known to be highly related to the accuracy of the annotation labels within the training dataset. Faults in the training dataset may lead to improper bias and network confusion. In this study, we evaluated the performance of the proposed architecture using two datasets. The first one is generated using a semi-automatic method detailed in \cite{Laakso2015BS}. The second (MICCAI 2012 dataset) is a commonly used dataset for the evaluation of similar approaches. However, with a comprehensive analysis of the golden truth segmentation of the second dataset, a notable limitation is observed. As shown in Fig.~\ref{datalimit0}, spike artifacts can be observed in region boundaries in axial and sagittal directions. However, more rigid boundaries can be found in the coronal direction. This may indicate that the manual segmentation is performed in the coronal direction and not corrected in other directions. These artifacts can mislead the assessment of segmentation accuracy as the golden truth may become inaccurate in axial and sagittal directions. To demonstrate this effect, we compare the segmentation of Caudate structure in different orientations in Fig.~\ref{datalimit}. In the first two rows, we demonstrate examples where both golden truth segmentation and SubForkNet are of high matching with reference to MRI. In the later rows, we show other examples where a strong mismatching is observed. However, it is clear that SubForkNet segmentation fits more with anatomical reference. It worth noting that accurate manual segmentation of deep brain regions is rather difficult and time-consuming task. Even with the above-discussed limitations, using MICCAI 2012 is preferable as standard dataset commonly used by the community for comparison.

\section{Conclusion}

In this study, a new end-to-end convolutional neural network architecture is proposed for the annotation of deep brain structures. A key feature of the proposed SubForkNet is the composition of single encoder sequence and individual decoders for different anatomical structure. Therefore, SubForkNet has more space to learn features associated to individual anatomical structures, which increase significantly the learning representation. Deep brain regions segmented by SubForkNet were embedded into personalized head models generated from ForkNet. Therefore, tDCS studies were conducted, and results indicated relatively high matching between the gold standard models and network segmented ones. These results suggest that the use of the convolutional neural network may take a leading role in personalized medicine, especially for clinical applications conducted through non-invasive brain electrostimulation.

Mathematica notebooks demonstrate the implementation of SubForkNet architectures and pre-trained networks are available for download at:

\href{http://github/erashed/Sub-ForkNet}{https://github.com/erashed/SubForkNet}

\section*{Acknowledgment}

This work was supported in part by JSPS Grant-in-Aid for Scientific Research (A), JSPS KAKENHI 17H00869. The authors would like to thank Mr. Akihiro Asai (Nagoya Institute of Technology, Nagoya, Japan) for his help with the tDCS montage setup.

\bibliography{Refs1}
\end{document}